\documentclass[]{spie}  

\usepackage{amsmath,amsfonts,amssymb}
\usepackage{graphicx}
\usepackage[colorlinks=true, allcolors=blue]{hyperref}
\usepackage{color}

\title{Optical Design of the TolTEC Millimeter-wave Camera} 

\author[ ]{Sean Bryan$^\mathrm{a}$\hspace{-1ex}}
\author[ ]{Jason Austermann$^\mathrm{b}$\hspace{-1ex}}
\author[ ]{Daniel Ferrusca$^\mathrm{c}$\hspace{-1ex}}
\author[ ]{Philip Mauskopf$^\mathrm{d}$\hspace{-1ex}}
\author[ ]{Jeff McMahon$^\mathrm{e}$\hspace{-1ex}}
\author[ ]{Alfredo Monta\~na$^\mathrm{c}$\hspace{-1ex}}
\author[ ]{Sara Simon$^\mathrm{e}$\hspace{-1ex}}
\author[ ]{Giles Novak$^\mathrm{f}$\hspace{-1ex}}
\author[ ]{David S\'anchez-Arg\"uelles$^\mathrm{c}$\hspace{-1ex}}
\author[ ]{Grant Wilson$^\mathrm{g}$,  for the TolTEC Team}

\affil[a]{School of Electrical, Computer \& Energy Engineering, Arizona State University, Tempe, AZ, USA}
\affil[b]{Quantum Sensors Group, National Institute of Standards and Technology, Boulder, CO, USA}
\affil[c]{Instituto Nacional de Astrof\'isica, \'Optica y Electr\'onica, Puebla, Mexico}
\affil[d]{School of Earth and Space Exploration, Arizona State University, Tempe, AZ, USA}
\affil[e]{Department of Physics, University of Michigan, Ann Arbor, MI, USA}
\affil[f]{Center for Interdisciplinary Exploration and Research in Astrophysics (CIERA) and Department of Physics \& Astronomy, Northwestern University, Evanston, IL, USA}
\affil[g]{Department of Astronomy, University of Massachusetts, Amherst, MA, USA}

\authorinfo{Send correspondence to S.A.B.: sean.a.bryan@asu.edu}
\begin{document}
\maketitle

\begin{abstract}
TolTEC is a new camera being built for the 50-meter Large Millimeter-wave Telescope (LMT) in Puebla, Mexico to survey distant galaxies and star-forming regions in the Milky Way. The optical design simultaneously couples the field of view onto focal planes at 150, 220, and 280 GHz. The optical design and detector properties, as well as a data-driven model of the atmospheric emission of the LMT site, inform the sensitivity model of the integrated instrument. This model is used to optimize the instrument design, and to calculate the mapping speed as an early forecast of the science reach of the instrument.
\end{abstract}



\section{INTRODUCTION}
\label{sec:intro}  

TolTEC on the LMT will open up a new view of the millimeter-wave sky. The instrument will discover and characterize distant galaxies by detecting the thermal emission of dust heated by starlight. The polarimetric capabilities of the camera will image dust interacting with magnetic fields in star-forming regions in the Milky Way. In addition to the galactic and extragalactic surveys already planned, TolTEC installed at the LMT will provide open observing time to the community. Other potential science goals include imaging galaxy clusters through the SZ effect and studying solar system objects. The project is reviewed more broadly in these proceedings \cite{wilson18}.

The optical design of the TolTEC uses mirrors, lenses, and dichroics to simultaneously couple a 4 arcminute diameter field of view onto three single-band focal planes at 150, 220, and 280 GHz. The optical design and detector properties, as well as a data-driven model of the atmospheric emission of the LMT site, inform an instrument sensitivity model we developed to aid in observing planning. We also used the model to optimize the instrument performance by selecting the edges of each passband to avoid atmospheric contamination, and also to calculate the loading onto the detectors to calculate the expected photon noise and optimize detector design. Opacity measurements at Sierra Negra, the LMT site, show that atmospheric conditions are excellent during the observing season in the driest months of the year with atmospheric opacity τ better than 0.06 for 25\% of that time and below than 0.10 for 50\% the same time\cite{ferrusca14,zeballos16}. These measurements at the site, along with MERRA satellite data, inform the atmospheric model used to calculate the instrument performance.

\section{Optical Design} 

TolTEC contains 6,300 polarization-sensitive detectors in three bands (150/220/280 GHz), simultaneously covering both linear polarizations and filling nearly the entire current field of view of the LMT in each band. To maximize sensitivity we adopted the conservative approach of using single-color detectors, dichroics, and single-color or reflective optics whenever possible. The system does use one two-color lens, which along with the single-color lenses is fabricated at the University of Michigan \cite{CMBS4-Technology-Book}. Using single-color focal planes allowed us to use the single-color LEKID detectors developed at NIST \cite{austermann18}. Single-color focal planes also allowed us to maximize the total instrument mapping speed by individually optimizing the detector spacing for each band to 1 $f\lambda$. In contrast, a hypothetical instrument using three-color detectors with spacing optimized for 220 GHz would be spatially undersampled at 280 GHz by $\sim 280/220$ = 27\%. This means that an optimized instrument could potentially have up to $\sim(280/220)^2$ = 62\% more 280 GHz mapping speed. (The aperture efficiency would also change somewhat, so the scaling is not perfectly linear with detector count.) A three-color focal plane would also be redundantly oversampled at 150 GHz by $\sim 220/150$ = 46\%. This means that due to spatial correlations in the radiation field, the effort spent fabricating and reading out the $\sim(220/150)^2$ = 115\% extra detectors would not translate into corresponding mapping speed gains. Designing a multi-color instrument around single-color or two-color detectors is thus more optimal in this regard.

\begin{figure}
\begin{center}
\begin{tabular}{c}
\includegraphics[width=0.95\textwidth]{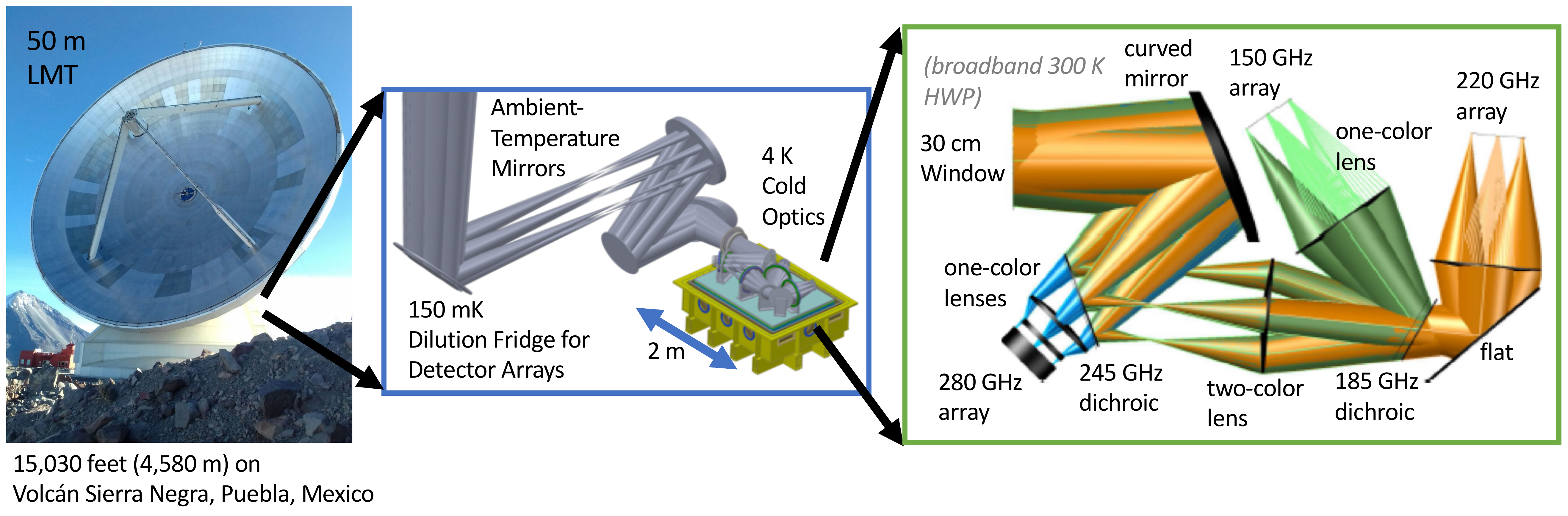}
\end{tabular}
\end{center}
\caption{Overview of TolTEC on the LMT. \label{lmt_toltec}}
\end{figure}

An overview of the TolTEC optical system on the LMT is shown in Figure~\ref{lmt_toltec}. Starting from the sky, the light is collected by the 50 m primary mirror and brought into the receiver cabin by the secondary mirror through a hole in the center of the primary. Four ambient temperature low-emissivity mirrors relay the beam into the cold optics of TolTEC. Passing through an ambient temperature half wave plate used for polarimetry, the beam enters the cold optical system through an UHMWPE window \cite{CMBS4-Technology-Book}. Inside the cryostat, after the Lyot stop light at all three TolTEC frequency bands is reflected and focused from a single curved mirror. The 280 GHz band is focused onto its detector array by two more single-color silicon lenses. The 220 GHz and 150 bands reflect from a 245 GHz dichroic, come to a focus, then pass through a two-color silicon lens for reimaging. The 220 GHz band reflects from a flat folding mirror and is focused by a single-color lens onto its detector array. The 150 GHz band reflects from a 185 GHz dichroic and is focused by a single-color silicon lens onto its detector array. Each of the three focal planes consists of an array of single-color polarizaton-sensitive LEKID detectors \cite{austermann18} coupled with feed horns and held at 100 mK by a dilution fridge.

\section{Modeling the Mapping Speed}

For optimizing the detector arrays and to assist in early observing planning, we developed a python code to calculate the expected mapping speed of TolTEC. The model includes a data-driven model of atmospheric emission and transmission at the site, modeling the telescope beam, detector noise and optical efficiency, the filter passbands, and an additional factor due to unremoved atmospheric fluctuations based on experience with the AzTEC instrument. We have made this code publicaly available on github at \url{github.com/TolTEC-Camera/MappingSpeedModel} for use by the community to model TolTEC or other instruments.

\subsection{Atmospheric Emission at the LMT Site}

The free publicly-available \textit{am}\footnote{\url{cfa.harvard.edu/\~spaine/am}} software can calculate absorption and emission from layers of gas made up of dry air, water vapor, and optionally other gases as well. It also smoothly interpolates between layers that have different gas pressure, as in layers of atmosphere at higher and higher elevations. It includes atmospheric profiles for astronomical sites such as Mauna Kea in Hawaii, the ALMA site in Chile, the South Pole, and recently the LMT site. These atmospheric profiles were created using archival weather satellite data compiled in the MERRA project run by NASA, covering different weather conditions at the site, at different times of year. Some sites have had these models validated with sounding balloon flights, and radiometer measurements. The MERRA data generally agrees with $\tau_{215\mathrm{GHz}}$ data taken at the LMT site\cite{ferrusca14,zeballos16}.

\begin{figure}
\begin{center}
\begin{tabular}{c}
\includegraphics[width=0.7\textwidth]{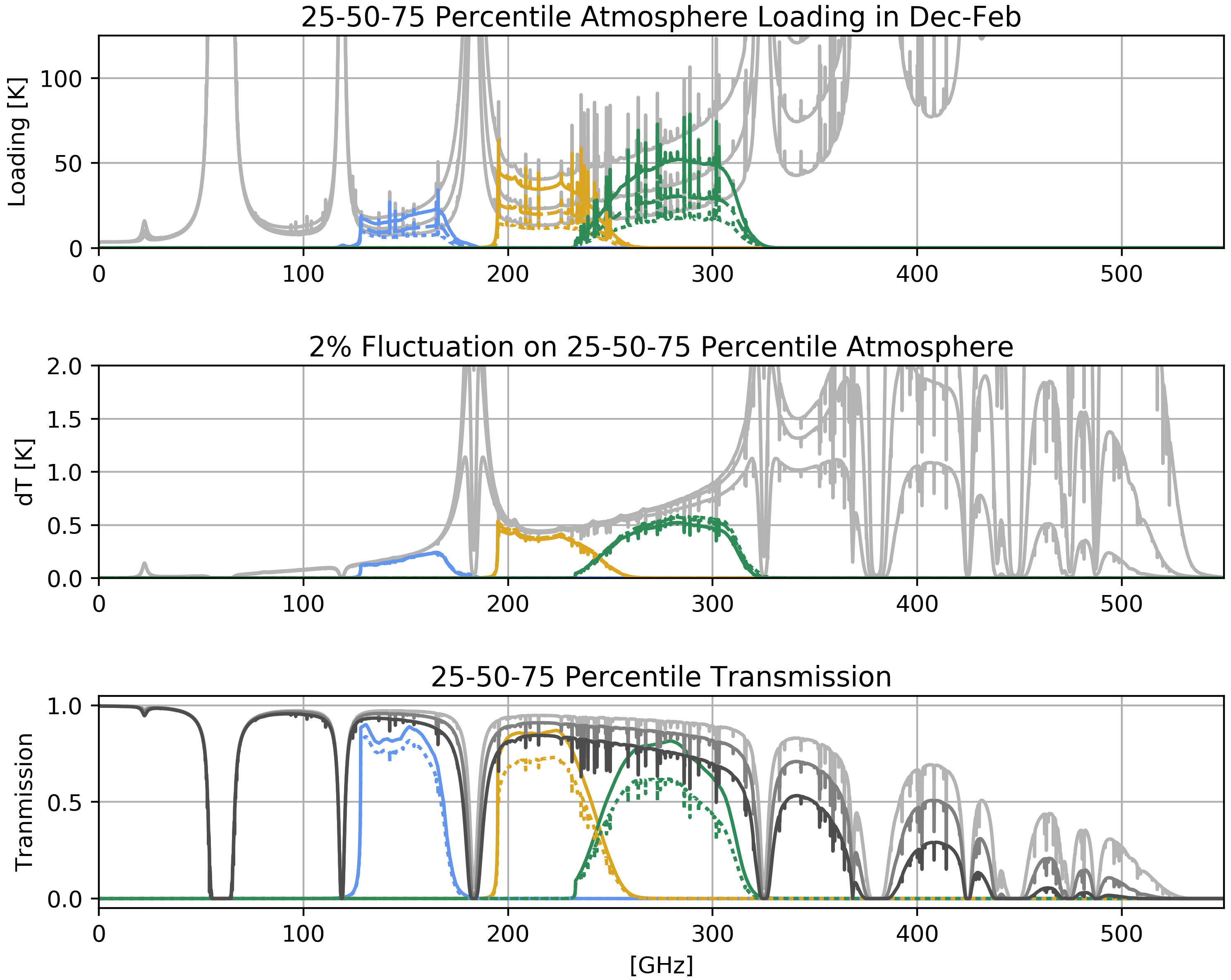}
\end{tabular}
\end{center}
\caption{Model of the LMT site from \textit{am} from December to February, with 25th, 50th, and 75th percentile weather conditions shown. The top panel shows the atmospheric loading, the middle panel shows the signal caused by atmospheric fluctuation, and the bottom panel shows atmospheric transmission. A model of the TolTEC filter passbands is overplotted, multiplied by the quantity shown in each panel. \label{am_model_results}}
\end{figure}

Simulated atmospheric transmission and emission from median weather conditions during the best time of the observing season, December to February, are shown in Figure~\ref{am_model_results}. Excellent windows exist at radio-50 GHz, 75-110 GHz, and the TolTEC bands of 128-170 GHz (blue), 195-245 GHz (yellow), and 245-310 GHz (green). For the very best weather days, it may be interesting to consider a future instrument with 335-360 GHz or even 395-415 GHz photometry bands, but photometry at frequencies higher than these may be challenging at this site. The \textit{am} code can also be run to forecast very narrow windows for heterodyne measurements at $\sim$THz frequencies at this site.

\subsection{Telescope Beam and Emission}

In general, unless the beam is aggressively truncated by an aperture stop, a telescope fed by a single-moded feed horn has a roughly constant beam size across its passband. Very roughly, that is because at the low end of the passband the beam of the feed horn at the focal plane is wider, which means it illuminates more of the primary mirror. On the high end of the passband, the feed horn has a narrower beam, illuminating less of the primary. A fully illuminated mirror at a long wavelength will have a similar beam to an under-illuminated mirror at a short wavelength. Because of this rough scaling argument, we make the approximation that the system has a single frequency-independent beam size for any given individual passband. For TolTEC, for the 150 GHz, 220 GHz, and 280 GHz bands, we assume 9.5'', 6.3'', and 5.0'' FWHM beams respectively. This does not figure into the detector noise calculation directly, but does matter for converting the detector noise into a mapping speed across a survey area.

The telescope as well as any warm mirrors will add thermal loading to the detectors. Here we assume that the LMT primary and secondary are at 273 K (i.e. near freezing temperature) and have a combined effective emissivity of 6\%, meaning they have an apparent antenna temperature of 16 K. For TolTEC, we also assume that we will have three ambient temperature polished mirrors at 290 K inside the receiver cabin (i.e. a slightly cool room temperature) each with an emissivity of 1\%, or a total antenna temperature of 9 K. This will all simply add to the antenna temperature of the atmosphere.

Thinking in broadcast mode, we assume that all of the detector sensitivity that makes it to the primary mirror will terminate on the cold sky. Because of surface roughness on the primary mirror, through Ruze scattering some of this sensitivity will not form the main beam, but instead will scatter. This efficiency of coupling to the main beam from a rough primary mirror is
\begin{equation}
TX_{telescope} = \exp{\left(-\left(\frac{4 \pi \sigma}{\lambda}\right)^2\right)},
\end{equation}
where $\sigma$ is the surface roughness. In-situ measurements of individual LMT surface segments show surface RMS variations that vary from 20 micron RMS to 40 micron RMS depending on the age and placement of the segment.  The project goal is to achieve a full system accuracy - including surface error, setting error, thermal and wind errors, and secondary mirror errors - of 76 microns RMS in 2019.  This corresponds to a combined Ruze scattering efficiency of 80\% for 150 GHz, 61\% for 220 GHz, 45\% for 280 GHz, and 38\% at the upper band edge of 310 GHz. For possible future photometric bands at higher frequencies, it would be 29\% at 350 GHz, and 19\% at 405 GHz.

\subsection{Loading and Detector Noise}

The \textit{am} code yields a calculated atmospheric emission temperature $T_{atm}$. After adding the thermal emission from the warm mirrors $T_{mirrors}$, this is then scaled by the instrument efficiency $TX_{inst}$ to yield the apparent antenna temperature at the detector (after the aperture efficiency, filters, feed horn, and detector optical efficiency factors). Note that the optical efficiency due to scattering in the telescope primary and secondary are not included here, since it is assumed that while the scattering does reduce the coupling from the detector into the main beam, we make the approximation that the scattered light terminates on the same atmospheric temperature as the main beam. This means that, at the detector, the apparent antenna temperature is
\begin{equation}\label{temp_scale}
T_{at-det} = \left( T_{atm} + T_{mirrors} \right) \times TX_{inst}.
\end{equation}
Assuming the loading has a Rayleigh-Jeans spectrum, the power at the single-polarization detector is
\begin{equation}\label{load}
P_0 = k T_{at-det} df
\end{equation}
where $k$ is Boltzmann's constant, and $df$ is the system bandwidth. Considering shot noise from the loading and wave noise from a single polarization,
\begin{eqnarray}\label{nep_both}
NEP_{shot} &=& \sqrt{2 k T_{at-det} h f df} \\
NEP_{wave} &=& \sqrt{2 P_0^2 / df},
\end{eqnarray}
where $f$ is the center frequency of the passband, and $h$ is Planck's constant. As discussed by Richards\cite{richards94}, near their Equation 11, the shot term can be roughly derived by noting that the fluctuations in photon number in one second are $<\Delta n^2> = kT / hf$. Taking the square root, multiplying by $h f$ to convert photon number to power, and then adding the $\sqrt{2}$ discussed later in Section~\ref{nefd_sec} to convert from seconds to bandwidth, we obtain the formula for shot noise. The $\sqrt{2}$ factor in the wave noise term is necessary in order for this expression to be consistent with the radiometer equation, which is the well-known wave-noise-limited sensitivity of any radiometer system.

Since the noise terms add in quadrature, and also assuming the detector has been optimized to have an internal noise level (due to thermal fluctuation noise in conventional bolometers, or generation-recombination noise in KIDs) that is a fraction $(1-X)$ of the photon noise level, the total at-detector NEP is:
\begin{equation}\label{scale_nep}
NEP_{tot}^{at-det} = \sqrt{X} \sqrt{NEP_{shot}^2 + NEP_{wave}^2}
\end{equation}
Here we do not review\cite{mauskopf18} the physics of internal detector noise.

Scaling the at-detector NEP up to the on-sky NEP requires scaling by the telescope+camera+detector system efficiency $TX_{sys}$, which must include the optical efficiency loss due to scattering on the primary mirror. It should also be scaled by atmospheric transmission $TX_{atmosphere}$.
\begin{equation}\label{nep_outside_atmosphere}
NEP = NEP_{tot}^{at-det} / (TX_{sys} TX_{atmosphere})
\end{equation}
This is the on-sky sensitivity of a single-detector with its single-polarization sensitivity.

\subsection{Point Source Detector Sensitivity}\label{nefd_sec}

The definition of Jansky flux density is $10^{-26}$ Watts of power, landing on 1 m$^2$ of telescope area, over 1 Hz of detector optical bandwidth, equally over both polarizations of light. The Noise Equivalent Flux Density (NEFD) is defined as the noise equivalent power in units of Jansky flux density, over an integration time of 1 second. Detector NEP is expressed in terms of W$\sqrt{\mathrm{Hz}}$, so since each Hz of bandwidth would only require half of a second of integration time to sample, a noise value expressed in $\sqrt{\mathrm{s}}$ is numerically smaller than a noise value expressed in $\sqrt{\mathrm{Hz}}$ by a factor of $\sqrt{2}$. These definitions all combine to the final formula for the NEFD of a polarized detector:
\begin{equation}\label{jansky}
NEFD = NEP \frac{2 \times 10^{26}}{\sqrt{2} \pi (D/2)^2 df},
\end{equation}
where the $2$ in the numerator is to account for the fact that the polarized detector only sees half the power coming from an unpolarized source, and $D$ is the primary mirror diameter. Note that the aperture efficiency due to the illumination of the primary by the feed horn was already treated in the previous calculation of the detector $NEP$, so an extra factor should not be added here.

\subsection{Arbitrary Passband}

To treat the frequency dependence of the detector passband, our code treats each frequency point $f_i$ in the atmospheric model spectrum as a ``virtual'' detector with a tophat passband $df = 10$ MHz being the frequency spacing between atmospheric model points. The $NEP(f_i)$ of each of these virtual detectors is calculated with its own optical efficiency (i.e. the detector response at that frequency), loading, atmospheric transmission, primary mirror scattering, and NEP to NEFD conversion. After repeating this calculation at each frequency bin, the total weighted average NEFD is formed by the weighted sum
\begin{equation}
NEFD = \sqrt{1 / \sum_{i} NEFD(f_i)^{-2}}.
\end{equation}
The at-detector NEP simply adds as the sum of squares, $NEP = \sum_{i} NEP(f_i)^2$.

\subsection{Instrument Mapping Speed}

\begin{table}[]
\centering
\caption{Calculated detector loading, individual detector noise levels, and total instrument mapping speed for TolTEC on the LMT for a median weather day from December to February.}
\label{noise_table}
\begin{tabular}{r|r|r|r|l}
                 & ~150 GHz~     & ~220 GHz~     & ~280 GHz~     &         \\ \hline
Passband                   & 128-170  & 195-245  & 245-310  & ~GHz        \\
$d \nu / \nu$ Bandwidth                  & 28\% & 23\% & 23\% &         \\ \hline
At-detector Loading        & 4.8 & 7.2 & 10.7 & ~pW      \\
At-detector NEP            & 50 & 72 & 95  & ~aW$/\sqrt{\mathrm{Hz}}$ \\
NET$_{\mathrm{CMB}}$       & 646 & 1979 & 4466& ~$\mu$K$\sqrt{\mathrm{s}}$   \\
NEFD                       & 0.59    & 1.01  & 1.46 & ~mJy$\sqrt{\mathrm{s}}$   \\ \hline \hline
Detector Count             & 900     & 1800        & 3600        &         \\
Total NET$_{\mathrm{CMB}}$ & 18.5  & 46.7 & 74.4 & ~$\mu$K$\sqrt{\mathrm{s}}$   \\
Total NEFD                 & 19.5 & 23.8 & 24.3 & ~$\mu$Jy$\sqrt{\mathrm{s}}$   \\
Mapping Speed              & 74.4  & 22.0 & 13.4 &   ~$\mathrm{deg}^2/\mathrm{mJy}^2/\mathrm{hour}$      \\ \hline \hline
Mapping Speed              & 10.5 & 3.1 & 1.9 &  ~$\mathrm{deg}^2/\mathrm{mJy}^2/\mathrm{hour}$       \\
\textit{Scaled by AzTEC Atm.} &             &             &             &        
\end{tabular}
\end{table}

Mapping speed is defined as the number of square degrees of sky that can be mapped to a 1 mJ noise level (i.e. a variance of 1 mJ$^2$) in beam-sized map pixels, in an hour. As such, it has units of square degrees, per mJ$^2$, per hour. Higher mapping speed implies a more sensitive camera that can map more sky at the desired map noise level (i.e. unlike detector noise, in this metric higher is better). In the previous sections, concluding in Equation~\ref{jansky}, we derived the sky-referenced noise of a single polarized detector in mJ$\sqrt{\mathrm{s}}$. This can be scaled to a mapping speed with the following considerations: With a noise level of $X$ mJ$\sqrt{\mathrm{s}}$, an RMS error of 1 mJ in a single beam would be achieved in $X^2$ seconds. Also, if a square degree contains $N$ beam-sized map pixels, it will take a total of $N \times X^2$ seconds to survey that area with a single detector to 1 mJ RMS. Defining the beam area as the total integrated solid angle of the equivalent Gaussian beam, the number of beam-equivalent-sized map pixels in a square degree $N_{beams}$ is
\begin{equation}\label{nbeams}
N_{beams} = \frac{(60\times60)^2}{FWHM_{arcsec}^2 \times \frac{\pi}{4 \ln 2}}.
\end{equation}
Scaling the mapping speed to hours requires dividing by the number of seconds in an hour, 3600. Combining this yields the mapping speed of a single detector:
\begin{equation}
\textrm{Single~Detector~Mapping~Speed} = \frac{3600}{N_{beams} \times X^2}~\mathrm{deg}^2/\mathrm{mJ}^2/\mathrm{hour}
\end{equation}
The instrument noise in mJ$\sqrt{\mathrm{s}}$ of a multi-detector camera scales with the number of detectors as $1/\sqrt{N_{det}}$. Since mapping speed is in mJ$^2$ variance, the variance scales as $1/N_{det}$, meaning that the total instrument mapping speed is
\begin{equation}\label{mapping-speed}
\textrm{Instrument~Mapping~Speed} = \frac{3600 \times N_{det}}{N_{beams} \times X^2}~\mathrm{deg}^2/\mathrm{mJ}^2/\mathrm{hour}
\end{equation}

This mapping speed assumes that detector white noise is the only relevant noise source in the system. Atmospheric fluctuations appear in detector timestreams in ground-based cameras at levels significantly above detector white noise at low audio frequencies. If these signals are common among multiple detectors, there are data analysis approaches for removing them from the data during mapmaking. Even after this removal, the overall white noise level is effectively increased, because some of the detector sensitivity was used to estimate these common mode signals and not used to estimate the sky maps. Empirically, from experience with the AzTEC camera at 280 GHz at the LMT, the atmosphere increased the effective white noise level by a factor of $\sqrt{7} = 2.65$, meaning that the mapping speed in mJ$^2$ dropped by a factor of 7. This should represent a worst-case scenario for building larger cameras in the future at this site. Modeling and experience suggests that atmospheric fluctuations have somewhat less impact at lower mm-wave frequencies. If the atmospheric fluctuations are common-mode across a small camera like AzTEC, but not common mode across larger angular scales, then this factor of 7 will continue to be the mapping speed penalty even for larger cameras. However, if the atmospheric signals turn out to be common mode across larger angular scales than the field of view of AzTEC, and common across our three frequencies, then in our larger camera more detectors can be used to estimate the atmospheric common mode. This would lead to a reduced noise penalty, so we currently believe that adopting a factor of 7 mapping speed penalty is a conservative way to forecast the expected science reach of the instrument for observation planning purposes.

Considering a specific survey area in square degrees, and a survey time in hours, the mapping speed can be used to calculate the beam-scale RMS of the map
\begin{equation}
\mathrm{Map~RMS} = \sqrt{\frac{\mathrm{Survey~Area}}{(\mathrm{Mapping~Speed}) \times (\mathrm{Survey~Time})}}.
\end{equation}
Calculated map RMS values for the Cloud to Cores, Fields in Filaments, Ultra-Deep Star Forming Galaxies, and the Wide Large-Scale Structure surveys are shown in Table~\ref{survey}.

\begin{table}[]
\centering
\caption{Calculated beam-scale map RMS values for the four defined 100-hour TolTEC surveys, assuming median conditions for the entire 4$\times$100 hours, and including the estimated reduced mapping speed due to atmospheric fluctuations.}
\label{survey}
\begin{tabular}{r|c|c|c|c|l}
 & \textit{Goal} &   ~150 GHz~          &     ~220 GHz~        &      ~280 GHz~       & ~Survey Area \\ \hline
Cloud to Cores~ & 0.240~mJ & 0.292~mJ & 0.538~mJ & 0.690~mJ & ~90 deg$^2$  \\
Fields in Filaments~ & 0.040~mJ    & 0.044~mJ & 0.080~mJ & 0.103~mJ & ~2 deg$^2$    \\
Ultra-Deep Star-Forming Galaxies~ & 0.025~mJ & 0.029~mJ & 0.054~mJ& 0.069~mJ & ~0.9 deg$^2$  \\
Wide LSS Survey~    & 0.260~mJ            & 0.308~mJ & 0.567~mJ & 0.727~mJ & ~100 deg$^2$ 
\end{tabular}
\end{table}

\section{Using and Verifying the Python Code}
We released the code to perform the calculations in this paper on github at \url{github.com/TolTEC-Camera/MappingSpeedModel}. Here we briefly discuss some of the details of the code, showing how it produces the tables and figures in this paper.

\subsection{Description of the Codes}

The atmospheric models were generated with \textit{am} using the LMT scenario files included in the latest version. An example of running the simulation is typing the following:

\verb|am LMT_DJF_50.amc 0 GHz 350 GHz 10 MHz 30 deg 1.0 > DJF_50.dat|
\\
The arguments are all documented inside the amc file itself, and are the minimum and maximum frequency, the simulation frequency resolution, the look angle relative to Zenith, and an overall scaling factor on how much water vapor is in the simulation. This particular scenario file is December-February, 50th percentile. Files for other times of year and other percentiles also are included in \textit{am}.

\verb|run_all_am_models.py| runs \textit{am} several times to make the atmosphere files shown in Figure~\ref{am_model_results} and used for the noise modeling.

\verb|waveguide_and_cardiff_bandpass.py| contains utility functions to load data about Cardiff filters, make calculations about waveguide cut-on and cutoffs, and to model the filter passbands for TolTEC.

\verb|make_filter_spectra_and_total_passbands.py| uses those utility functions to make Figure~\ref{am_model_results}. It also makes the file \verb|model_passbands.npz| which is a NumPy file storing the passband models used later in the noise modeling.

\verb|toltec_noise_const.py| reproduces the results derived by hand in Section~\ref{toy_model}.

\verb|toltec_noise.py| is the full TolTEC noise model, producing the results in Tables~\ref{noise_table} and \ref{survey}.

\subsection{Toy Model for Code Testing} \label{toy_model}

Here we use the code to model a simple idealized instrument, and confirm that the code yields the same mapping speed as a calculation by hand. We consider an ideal tophat band from 245-310 GHz, 70$\%$ detector optical efficiency, 35$\%$ aperture efficiency for the feed horns illuminating the primary mirror, 47$\%$ primary mirror optical efficiency, and for the moment ignoring the emission of the warm optical elements. As a rough and fairly optimistic approximation of the atmosphere, assume 20 K loading from the sky, and 92$\%$ atmospheric transmission.

Going through the calculation by hand, Equation~\ref{temp_scale} shows that after instrument transmission, the antenna temperature at the detector is 20 K$/(0.70\times0.35)$ = 4.9 K. This means that the loading in Equation~\ref{load} is 4.4 pW. Using Equation~\ref{nep_both} yields a shot noise of 40 aW$\sqrt{\mathrm{Hz}}$ and a wave noise of 24 aW$\sqrt{\mathrm{Hz}}$. Equation~\ref{scale_nep} gives the total noise as 47 aW$\sqrt{\mathrm{Hz}}$ at the detector. Scaling this back out to the sky with Equation~\ref{nep_outside_atmosphere}, and converting to Janskies with Equation~\ref{jansky}, yields a single-detector NEFD of 0.53 mJ$\sqrt{\mathrm{s}}$. Running the full python code \verb|toltec_noise_const.py| reproduces these numbers, verifying the code.

\subsection{Full TolTEC Model}

Running the code \verb|toltec_noise.py| uses the same code tested in the previous subsection, but to model the full TolTEC instrument. It yields the results shown in Tables~\ref{noise_table} and \ref{survey}.

\section{Conclusions}

TolTEC is currently under construction and is scheduled for deployment onto the LMT soon. Its unprecedented mapping speed, angular resolution, and polarimetric capabilities at 150, 220, and 280 GHz will transform millimeter-wave astronomy. The forecast sensitivity presented here will be useful for the community to plan how to best use TolTEC to its fullest scientific potential.

\acknowledgments     
 
TolTEC is supported by NSF grant AST-1636621.  


\bibliography{report}   

\begin{thebibliography}{1}

\bibitem{wilson18}
Wilson, G. and {the TolTEC Team}, ``The toltec project: a millimeter wavelength
  imaging polarimeter,'' {\em Submitted to Proc. SPIE}  (2018).

\bibitem{ferrusca14}
Ferrusca, D. and Contreras, J.~R., ``Weather monitor station and 225 ghz
  radiometer system installed at sierra negra: the large millimeter telescope
  site,'' {\em Proc. SPIE}~{\bf 9147},  914730 (2014).

\bibitem{zeballos16}
Zeballos, M., Ferrusca, D., Contreras, J.~R., and Hughes, D., ``Reporting the
  first 3 years of 225-ghz opacity measurements at the site of the large
  millimeter telescope alfonso serrano,'' {\em Proc. SPIE}~{\bf 9906},  99064U
  (2016).

\bibitem{CMBS4-Technology-Book}
{CMB-S4 Collaboration}, ``{CMB-S4 Technology Book},'' (2017).

\bibitem{austermann18}
Austermann, J., Beall, J., Bryan, S., Dober, B., Gao, J., Hilton, G., Hubmayr,
  J., Mauskopf, P., McKenney, C., Simon, S., Ullom, J., Vissers, M., and
  Wilson, G., ``Millimeter-wave polarimeters using kinetic inductance detectors
  for toltec and beyond,'' {\em Submitted to JLTP}  (2018).

\bibitem{richards94}
Richards, P.~L., ``Bolometers for infrared and millimeter waves,'' {\em Journal
  of Applied Physics}~{\bf 76}(1),  1--24 (1994).

\bibitem{mauskopf18}
Mauskopf, P.~D., ``Transition edge sensors and kinetic inductance detectors in
  astronomical instruments,'' {\em Publications of the Astronomical Society of
  the Pacific}~{\bf 130}(990),  082001 (2018).

\end{thebibliography}
\bibliographystyle{spiebib}   

\end{document}